\documentclass[%
pof,
amsmath,
amssymb,
prl,
floats,
]{revtex4}
\usepackage{graphicx}
\usepackage{dcolumn}
\usepackage{bm}
\usepackage{natbib}
\usepackage{amsfonts}
\usepackage{subfigure}
\begin{document}
\title{A streamwise-constant model of turbulent pipe flow}
\author{Jean-Loup Bourguignon}
\author{Beverley J. McKeon}
\affiliation{Graduate Aerospace Laboratories, California Institute of
  Technology, Pasadena, CA 91125, USA}

\begin{abstract}
A streamwise-constant model is presented to investigate the basic mechanisms responsible for the change in mean flow occuring during pipe flow transition. Using a single forced momentum balance equation, we show that the shape of the velocity profile is robust to changes in the forcing profile and that both linear non-normal and nonlinear effects are required to capture the change in mean flow associated with transition to turbulence. The particularly simple form of the model allows for the study of the momentum transfer directly by inspection of the equations. The distribution of the high- and low-speed streaks over the cross-section of the pipe produced by our model is remarkably similar to one observed in the velocity field near the trailing edge of the puff structures present in pipe flow transition.
Under stochastic forcing, the model exhibits a quasi-periodic self-sustaining cycle
characterized by the creation and subsequent decay of ``streamwise-constant puffs'', so-called due to the good agreement between the temporal evolution of their velocity field and the projection of the velocity field associated with three-dimensional puffs in a frame of reference moving at the bulk velocity. 
We establish that the flow dynamics are
relatively insensitive to the regeneration mechanisms invoked to produce near-wall streamwise vortices and that using small, unstructured background
disturbances to regenerate the streamwise vortices is sufficient to capture the formation of the high- and low-speed streaks and their segregation leading to the blunting of the velocity profile characteristic of turbulent pipe flow.
\end{abstract}
\maketitle
\section{Introduction}
Linear models of pipe flow, e.g. \cite{Schmid}, capture the general characteristics of the coherent structures present in the flow but are unable to reproduce the change in mean flow associated with transition to turbulence. Linear studies showed that linear non-normality is required to sustain turbulence \citep{Henningson} and so are the terms linear in turbulence fluctuations \citep{KimLim}. The non-normality of the linear Navier-Stokes (LNS) operator results in large amplification of disturbances and transient growth of initial perturbations. However, the transient growth of the most amplified structures modify the mean flow in a way that reduces the amplification potential, i.e. the non-normality \citep{Waleffe}. Hence, realistic models for pipe flow transition must include mechanisms leading to a change in mean flow as the perturbations develop.\\
\\
In a first attempt to obtain sustained large amplification of disturbances, \cite{Bagget} introduced a mostly linear model for pipe flow transition containing a generic nonlinearity used to bootstrap the perturbations. However, the generic nonlinearity is unphysical and does not lead to the characteristic blunting of the velocity profile associated with turbulent flow. Since the blunting of the velocity profile can be accurately captured by the full Navier-Stokes equations (NSE), a more convenient approach would be to go back to the NSE and invoke some simplifying assumptions in order to reduce the complexity of the model instead of starting from the LNS and adding some sort of nonlinearity. For example, the assumption of streamwise invariance leads to a model containing only 2 partial differential equations instead of 4 for the NSE. Streamwise-constant models describe the evolution of the 3 components of velocity in a plane perpendicular to the mean flow and are equivalently referred to as 2D/3C. The unforced 2D/3C model for Couette flow was shown by \cite{Antonis} to be globally stable for all Reynolds numbers, i.e. it has a unique fixed point corresponding to the laminar flow. A stochastically forced 2D/3C model was used by \cite{Dennice} to study Couette flow and successfully captured the blunting of the velocity profile. The model also generated structures similar to the streamwise-elongated vortices and streaks observed in experiments. The stochastically forced 2D/3C model exploits the large amplification of background disturbances due to the non-normality of the linearized operator \citep{Ioannou} which is maximum for streamwise-constant disturbances \citep{Bamieh}. The latter authors showed that streamwise-constant disturbances are amplified proportionally to $Re^3$ vs. $Re^{3/2}$ for streamwise-variant disturbances. In addition, \cite{Jovanovic} demonstrated that the largest amplification is only obtained by forcing in the plane perpendicular to the mean flow and is observed in the streamwise velocity component.\\
\\
Pipe flow is also well suited for an assumption of streamwise invariance since streamwise-elongated coherent structures have been shown to play an important role during transition, e.g. \cite{Eckhardt2}, as well as in fully developed turbulence, e.g. \cite{Kim2}, \cite{Guala}, \cite{Hutchins07}, and \cite{Marusic10}. A streamwise-constant model for pipe flow was derived by \cite{Antonis} and shown by \cite{Bobba} to be globally stable for all Reynolds numbers. The streamwise-elongated coherent structures in pipe flow take the form of quasi-streamwise vortices and streaks of streamwise velocity. The most amplified mode of the LNS, based on an energy norm, is indeed streamwise-constant with an azimuthal wavenumber $n=1$ and features a pair of counter-rotating vortices which create streaks by convecting streamwise momentum \citep{Schmid}.
\cite{Reshotko} studied the spatial evolution of optimal disturbances in pipe flow in contrast to previous studies focusing on the temporal evolution and argued that a spatial study is better suited for comparison with experiments in which the disturbances are growing as they convect downstream. Reshotko \textit{et al.} concluded that the most amplified disturbances are stationary and have an azimuthal wavenumber $n=1$. The nonlinear study by \cite{Ioannou} of turbulent Couette flow also emphasizes the dominant role played by the streamwise-constant modes in the flow dynamics. Based on an energy transfer analysis, the latter authors showed that the modes $(0,\pm n)$ dominate energy extraction from the laminar base flow using linear non-normal mechanisms, and maintain the mean turbulent flow via their nonlinear interaction. Note that the mean turbulent mode $(0,0)$ does not extract energy directly from the laminar base flow.\\
\\
The laminar base flow becomes unstable when streamwise-constant vortices and streaks are superposed due to the creation of inflection points \citep{Meseguer} which sustain the growth of infinitesimal $3D$ disturbances until the streaks decay \citep{Zikanov}. \cite{Waleffe} argued that the $3D$ disturbances can regenerate the rolls by nonlinear interaction which consequently create the streaks by convecting streamwise momentum, leading to a self-sustaining process (SSP). The regeneration mechanisms invoked by \cite{Waleffe} were later revisited by \cite{Schoppa} who introduced a new mechanism called streak transient growth to regenerate the rolls based on the instability of the streaks. The $3D$ infinitesimal perturbations exhibit transient growth and evolve into sheets of streamwise vorticity which are then stretched by the mean shear and collapse, resulting in the formation of streamwise rolls. The SSP dominates the near-wall cycle in fully developed turbulence and plays an important role in pipe flow transition by maintaining the puffs, see \cite{vanDoorne}.\\ 
\\
Traditionally, transition to turbulence in pipe flow has been characterized by the creation of puffs and slugs \citep{Champagne}. Puffs are the flow response to large amplitude disturbances at low Reynolds number, e.g. $Re \approx 2000$, and are characterized by a sharp trailing edge and a smooth leading edge whereas slugs are created by low amplitude disturbances at larger Reynolds number, $Re>3000$ and have sharp leading and trailing edges \citep{Champagne}. The puffs are sustained via a SSP taking place near the trailing edge and which is characterized by the creation of low-speed streaks inside the puff which convect slower than the puff and create a shear layer at the boundary with the laminar flow at the back of the puff \citep{Shimizu}. The shear layer is subject to Kelvin-Helmholtz instability resulting in the creation of streamwise vortices by roll-up of vortex sheets. The streamwise vortices propagate faster than the puff and maintain the turbulence inside the puff as they re-enter it. The quasi-periodic generation of streamwise vortices near the trailing edge of the puffs, where the transition from laminar to turbulence takes place, was also reported by \cite{vanDoorne}. \cite{Hof2} suggest a new driving mechanism for puffs based on the formation of inflection points in the velocity profile near the trailing edge of the puff whose instability sustains turbulence inside the puff. Features of the SSP were also observed in travelling waves, see \cite{vanDoorne}. \\
\\
The clear distinction between puffs and slugs made by \cite{Champagne} was later questioned by \cite{Darbyshire} who observed mixed occurences of puffs and slugs. More recently, \cite{Duguet} argued that slugs are out-of-equilibrium puffs and therefore cannot exist together with stable equilibrium puffs. Stable equilibrium puffs are observed at $Re \approx 2200$ and convect slightly slower than the mean flow. Equilibrium puffs keep a constant length as they travel downstream and are separated by regions of laminar flow which are necessary to sustain them, as noticed by \cite{Lindgren}, see also \cite{Hof2}. Equilibrium puffs represent a minimal flow unit able to sustain turbulence. The particle-image-velocimetry (PIV) measurements of \cite{Hof} show that the dominant flow structures inside a puff are quasi-streamwise vortices and streaks and are independent of the method used to generate the puff \citep{Champagne}. \cite{Hof} highlight the similarity between the travelling wave (TW) solutions of the NSE and the velocity field near the trailing edge of a puff. 
At larger Reynolds number, the puffs expand as they convect downstream and tend to merge together. The puffs become unstable via a Kelvin-Helmholtz instability of the wall-attached shear layers \citep{Duguet}, resulting in the formation of slugs which keep expanding until the whole flow domain is turbulent.\\
\\
In this paper, we present a streamwise-constant model for turbulent pipe flow which allows us to isolate the
basic mechanisms governing the dynamics that
result in the blunting of the velocity profile and generates streamwise-constant versions of a puff. The model is described in the next section, together with the numerical methods employed to simulate the flow. In the third section, we use a deterministic forcing to show that the blunting is due to the segregation of the low-speed streaks from the
high-speed streaks, defined with respect to the laminar flow, by the
nonlinear coupling terms between the in-plane and axial velocities. In the fourth section, we describe the response of our model to stochastic forcing in the cross-stream plane. We conclude this study by summarizing the main achievements obtained with the 2D/3C model for pipe flow.
\section{Description of the model and numerical methods}
\begin{figure}[!h]
\centering
    \includegraphics[scale=0.25]{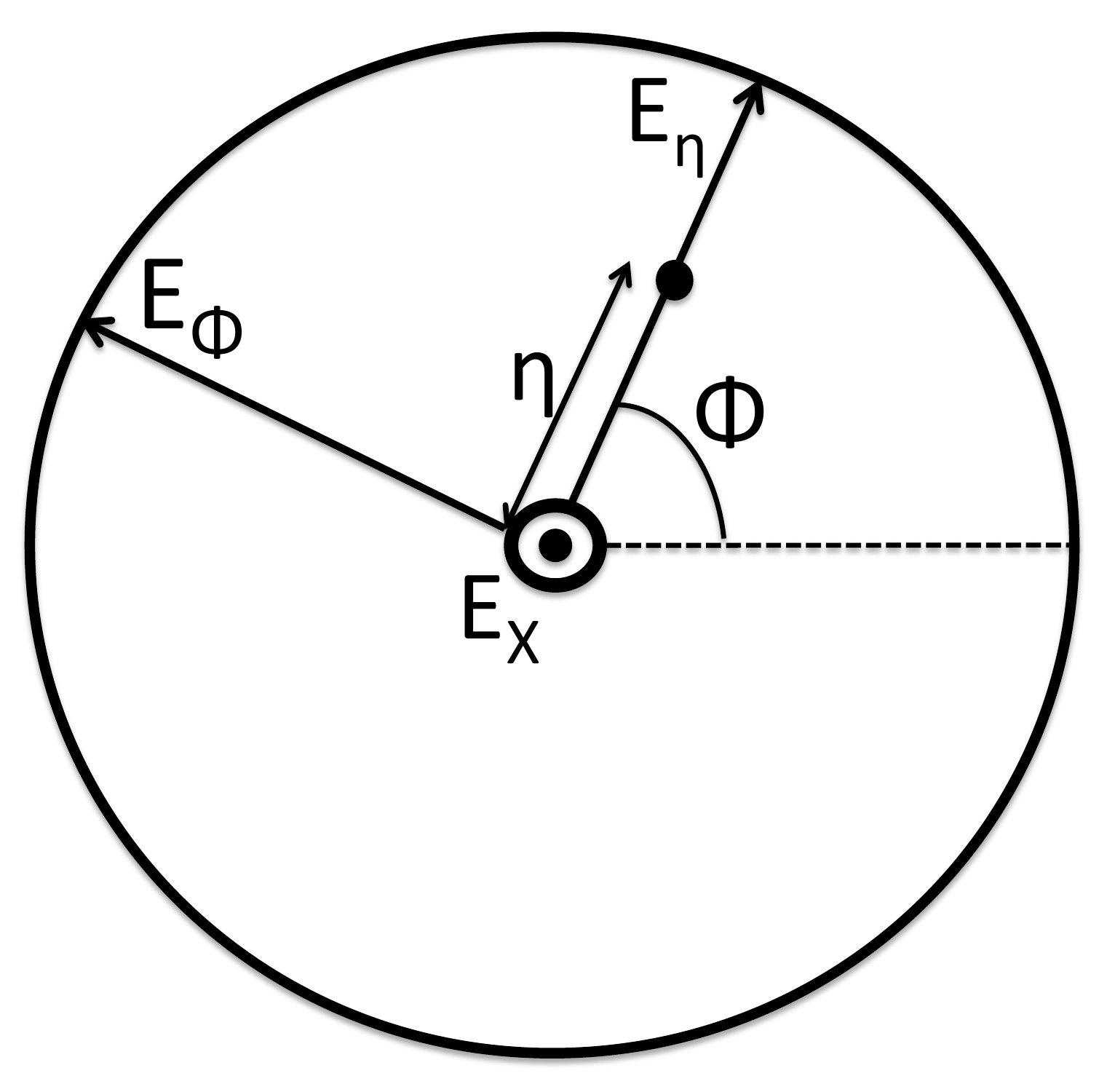}
\caption{The coordinate system
used to project the Navier-Stokes equations.
 \label{Coord}}
\end{figure}

The streamwise-constant model of turbulent pipe flow is derived from the NSE written in cylindrical coordinates under the assumption of streamwise invariance, i.e. it constitutes a projection of the NSE onto the streamwise direction. The NSE are nondimensionalized based on the pipe radius $R$ and the bulk velocity $\bar{U}$, i.e. $\eta=\frac{r}{R}$, $\tau=\frac{\bar{U} t}{R}$ and $Re=\frac{2 R \bar{U}}{\nu}$, where $\nu$ is the kinematic viscosity. Continuity is enforced via the introduction of a dimensionless streamfunction $\Psi$ whose evolution equation is obtained by taking the curl of the NSE projected in the axial direction. The model consists of an evolution equation for the streamfunction, from which the radial and azimuthal velocities can be derived, and an evolution equation for the axial velocity in terms of the deviation from the laminar profile corresponding to the axial momentum balance. The deviation of the axial velocity from laminar illustrates how the flow evolves away from the laminar state and is defined as $u_{x}=\tilde{u}_{x}-U$, where $\tilde{u}_{x}$ is the instantaneous axial velocity and $U$ is the laminar base flow. The 2D/3C model is written
\[\begin{cases}
\frac{\partial \Delta \Psi}{\partial \tau}=\frac{2}{Re} \Delta^2
\Psi+\mathcal{N}_{\psi},\\
\frac{\partial u_{x}}{\partial
  \tau}
=C-\frac{1}{\eta}\frac{\partial
  \Psi}{\partial \phi}\frac{\partial U}{\partial
  \eta}-\frac{1}{\eta}\frac{\partial \Psi}{\partial \phi}\frac{\partial
  u_{x}}{\partial \eta}+\frac{1}{\eta}\frac{\partial \Psi}{\partial
  \eta}\frac{\partial u_{x}}{\partial \phi}+\frac{2}{Re} \Delta u_{x},\\
\end{cases}
\]
with no-slip and no-penetration boundary conditions on the wall of the pipe. The model was derived in \cite{Antonis} and the coordinate system used to project the NSE is represented on Figure \ref{Coord}. 
Only the streamfunction equation is forced, based on the study of \cite{Jovanovic} which showed that maximum amplification is obtained by forcing in the cross-sectional plane. The small-amplitude zero-mean forcing $\mathcal{N}_{\psi}$ represents the perturbations that are always present in experiments, e.g. wall roughness, vibrations, non-alignment of the different sections of the pipe, thermal effects, and takes into account the effects not modeled by the streamwise invariance approximation. The nonlinear terms in the governing equation for the streamfunction are neglected in order to obtain to simplest model able to capture the blunting of the velocity profile and also because their effects can be incorporated into the unstructured forcing term $\mathcal{N}_{\psi}$. Moreover the study of \cite{Dennice} showed that there are no significant differences in the Couette flow statistics obtained from the model based on a  linearized streamfunction equation compared to the fully nonlinear 2D/3C model. The bulk velocity is maintained constant by adjusting the pressure gradient $C$, i.e. the Reynolds number is held constant for each study, and the radial and azimuthal velocities are defined by $u_{r}=\frac{1}{\eta}\frac{\partial \Psi}{\partial \phi}$, $u_{\phi}=-\frac{\partial \Psi}{\partial \eta}$. The streamwise velocity behaves as a passive scalar convected by the in-plane velocities.\\
\\
The 2D/3C model is discretized using a spectral-collocation method based on Chebyshev polynomials in the radial direction and Fourier modes in the azimuthal direction, associated with a third-order semi-implicit time stepping scheme described in \cite{Spalart}. The singularity at the origin of the polar coordinate system is avoided by re-defining the radius from $-1$ to $1$ and using an even number of grid points in the radial direction, as in \cite{Heinrichs}. Three Sylvester equations are written respectively for $\Delta\Psi$, $\Psi$, and $u_{x}$, associated with homogeneous Dirichlet boundary conditions, and are solved using an optimized Sylvester equation solver from the \textsc{slicot} numerical library \citep{slicot}. The boundary conditions (BCs) $\Psi=0$ and $u_{x}=0$ at the wall correspond respectively to no-penetration and no-slip in the axial direction. The no-slip BC in the azimuthal direction is enforced by adding particular solutions to the streamfunction, following the influence matrix method for linear equations, see \cite{Peyret}, such that the azimuthal velocity $u_{\phi}=-\frac{\partial \Psi}{\partial \eta}$ vanishes at the wall.\\
\section{Simplified 2D/3C model with deterministic forcing}
\begin{figure}[!h]
\centering
\subfigure[] 
{
    \label{fig:sub:a}
\includegraphics[scale=0.35]{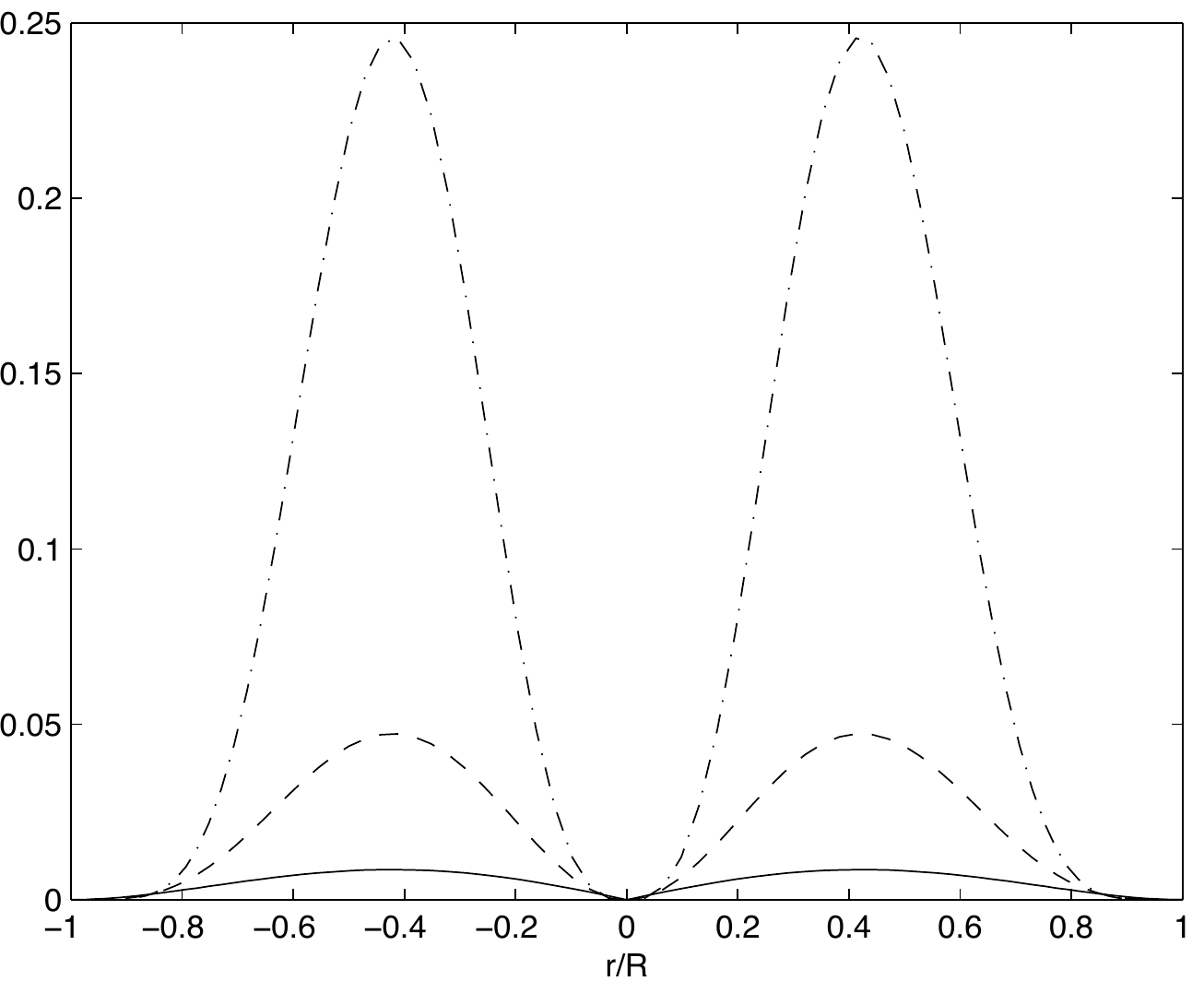}
}
\subfigure[]
{
    \label{fig:sub:b}
\includegraphics[scale=0.35]{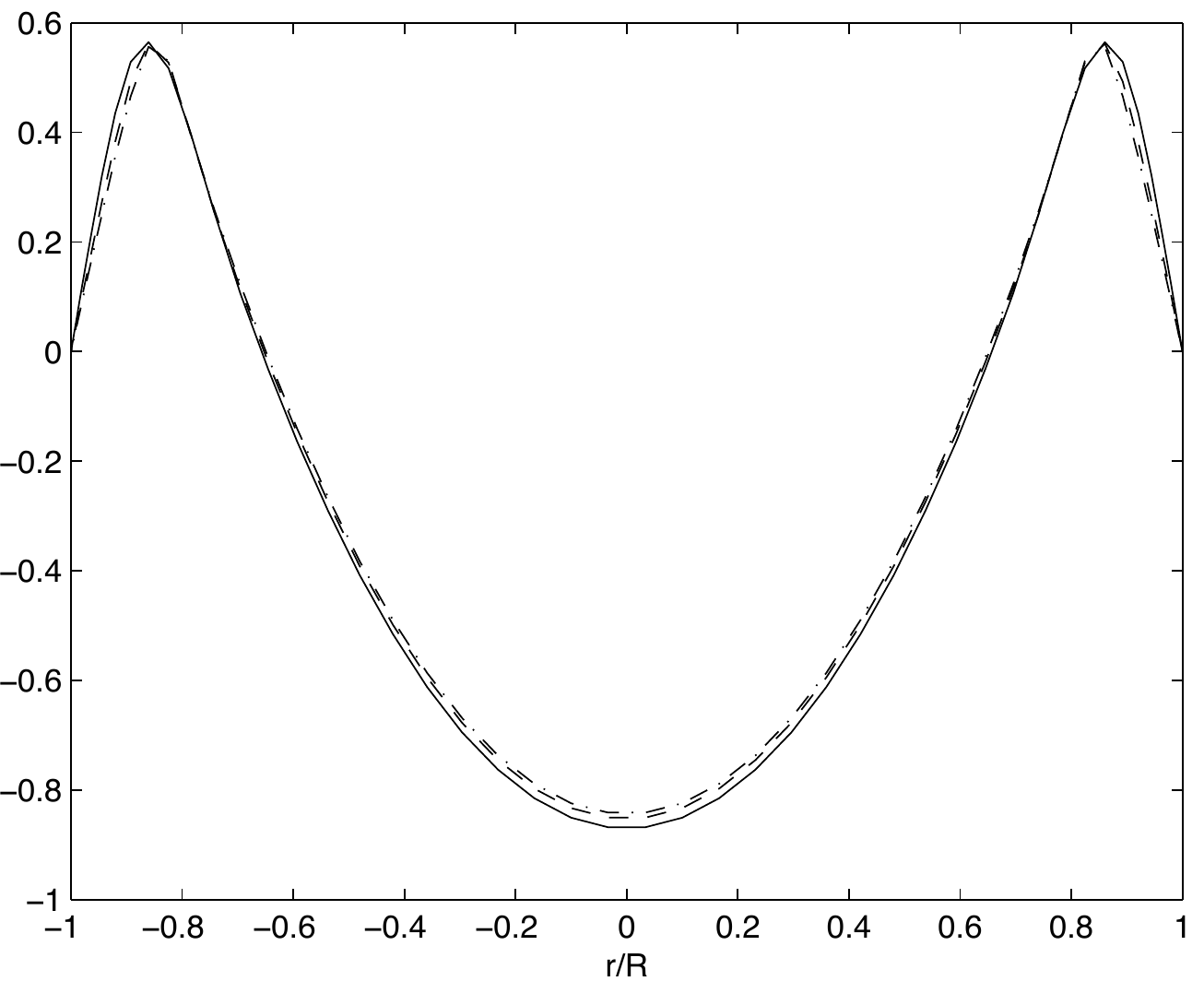}
}
\caption{(a) Streamfunctions $\Psi_{1}(\eta)$ and (b) corresponding velocity profiles $u_{0}(\eta)$ for $\Psi_{1,a}(\eta)=0.033(\eta-3\eta^3+2\eta^4)$ (solid), 
$\Psi_{1,b}(\eta)=0.7(\eta-3\eta^3+2\eta^4)^2$ (dashed), and $\Psi_{1,c}(\eta)=14(\eta-3\eta^3+2\eta^4)^3$ (dash-dot).
 \label{Analytic}}
\end{figure}

\begin{figure}[!h]
\centering
\subfigure[] 
{
    \label{fig:sub:a}
    \includegraphics[scale=0.35]{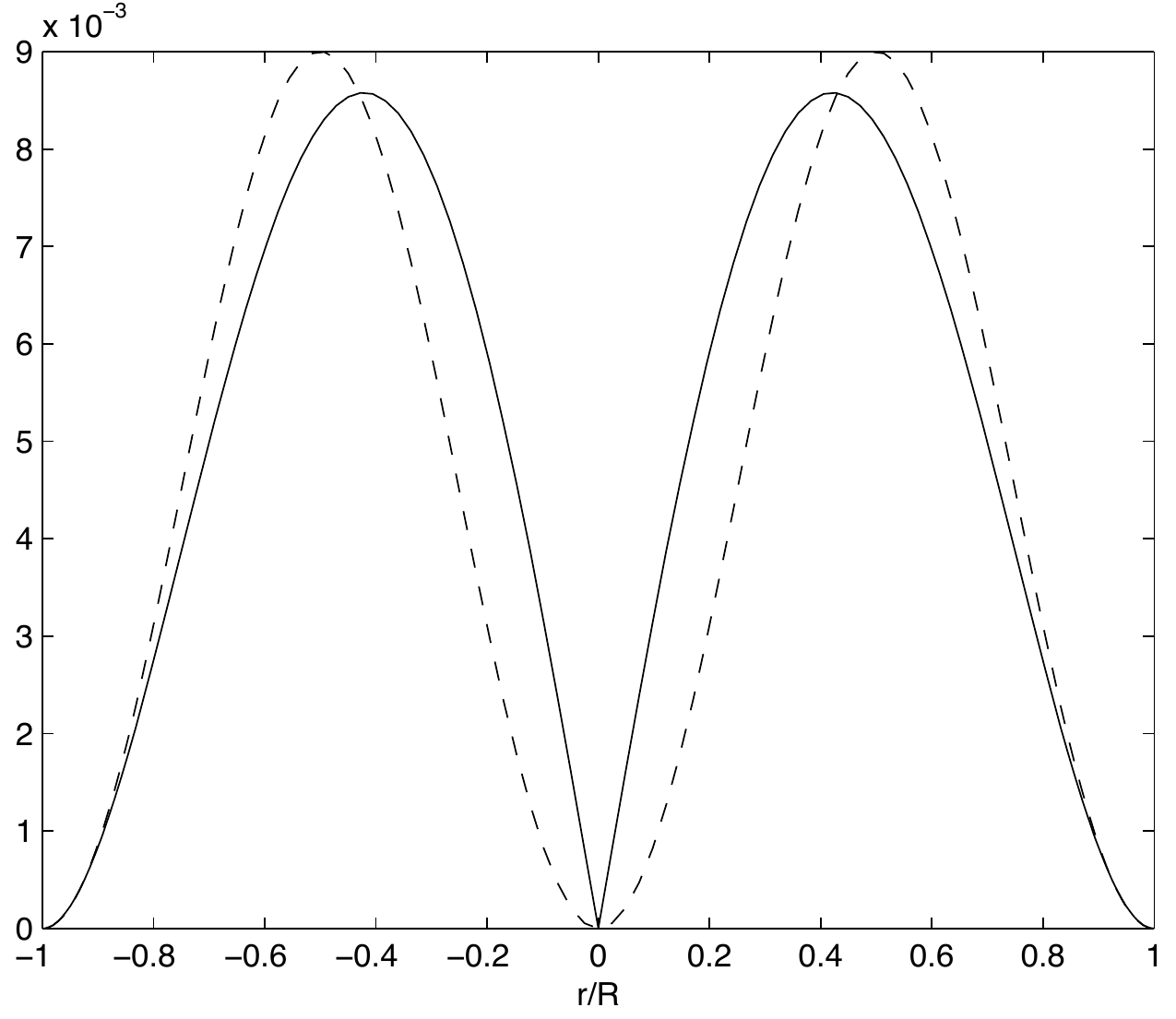}
}
\subfigure[]
{
    \label{fig:sub:b}
    \includegraphics[scale=0.35]{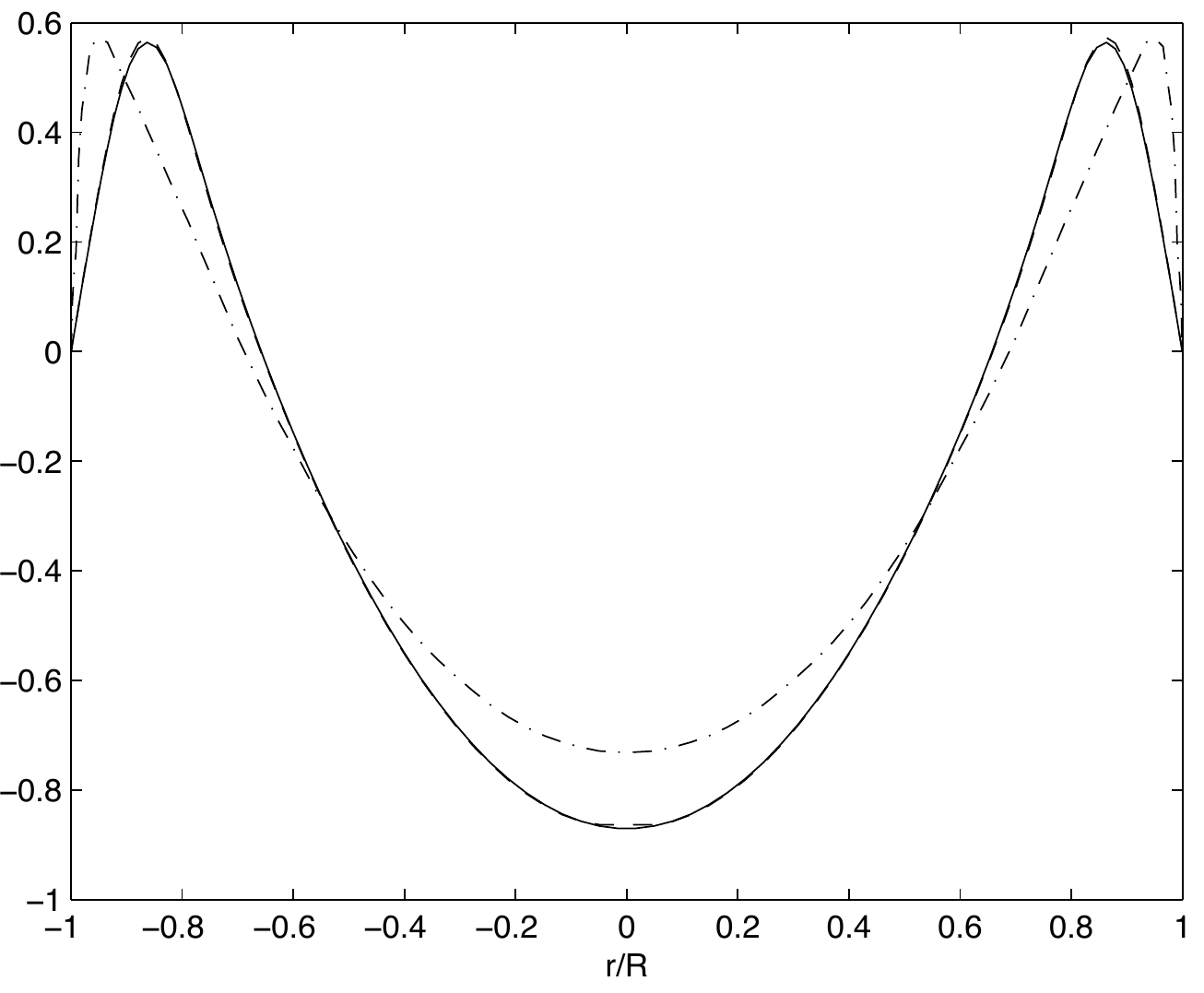}
}

\caption{(a) Streamfunctions $\Psi_{1}(\eta)$ and (b) corresponding velocity profiles $u_{0}(\eta)$ for $\Psi_{1,a}(\eta)=0.033(\eta-3\eta^3+2\eta^4)$ (solid), 
$\Psi_{1,b}(\eta)=0.009\sin(\pi \eta)^2$ (dashed), and experimental
velocity profile of \cite{Toonder} at Re=24,600 (dash-dot). Note that the solid and dotted lines for the velocity profiles are identical, except at their extrema where they differ slightly, i.e. by less than 0.5\%.
 \label{Compa}}
\end{figure}

\begin{figure}[!h]
\centering
\subfigure[] 
{
    \label{fig:sub:a}
    \includegraphics[scale=0.29]{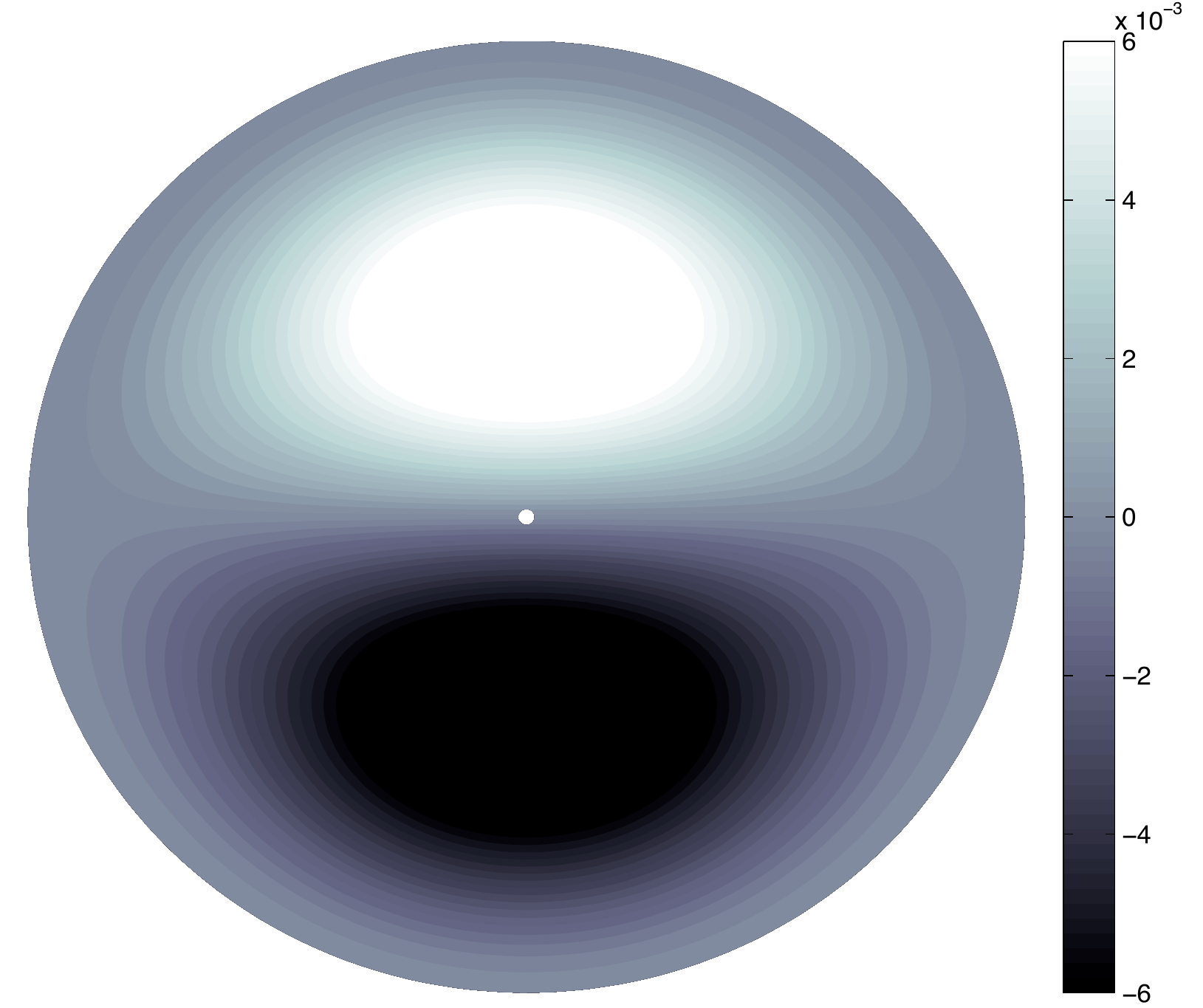}
}
\subfigure[]
{
    \label{fig:sub:b}
    \includegraphics[scale=0.34]{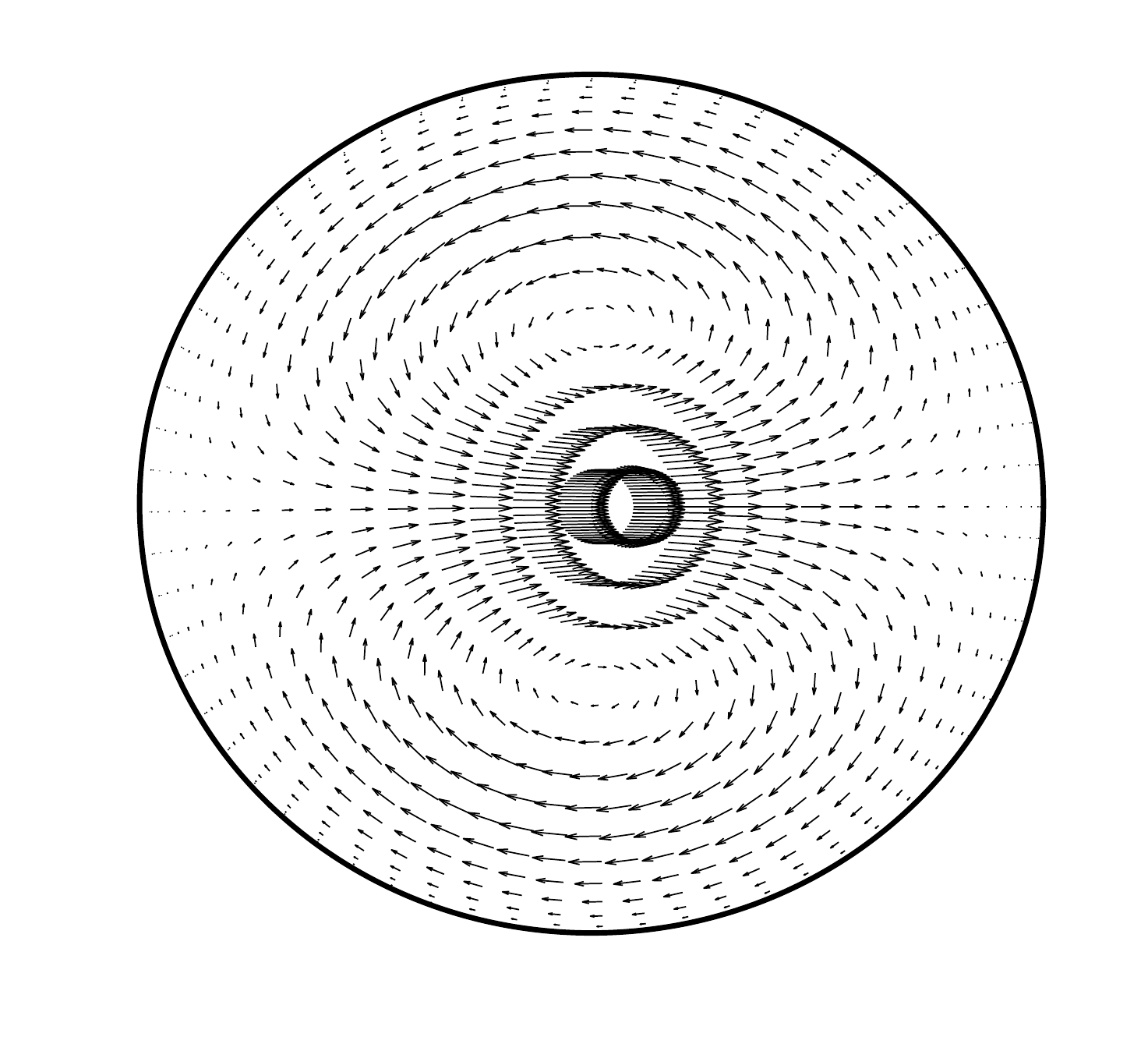}
}
\subfigure[]
{
    \label{fig:sub:c}
    \includegraphics[scale=0.29]{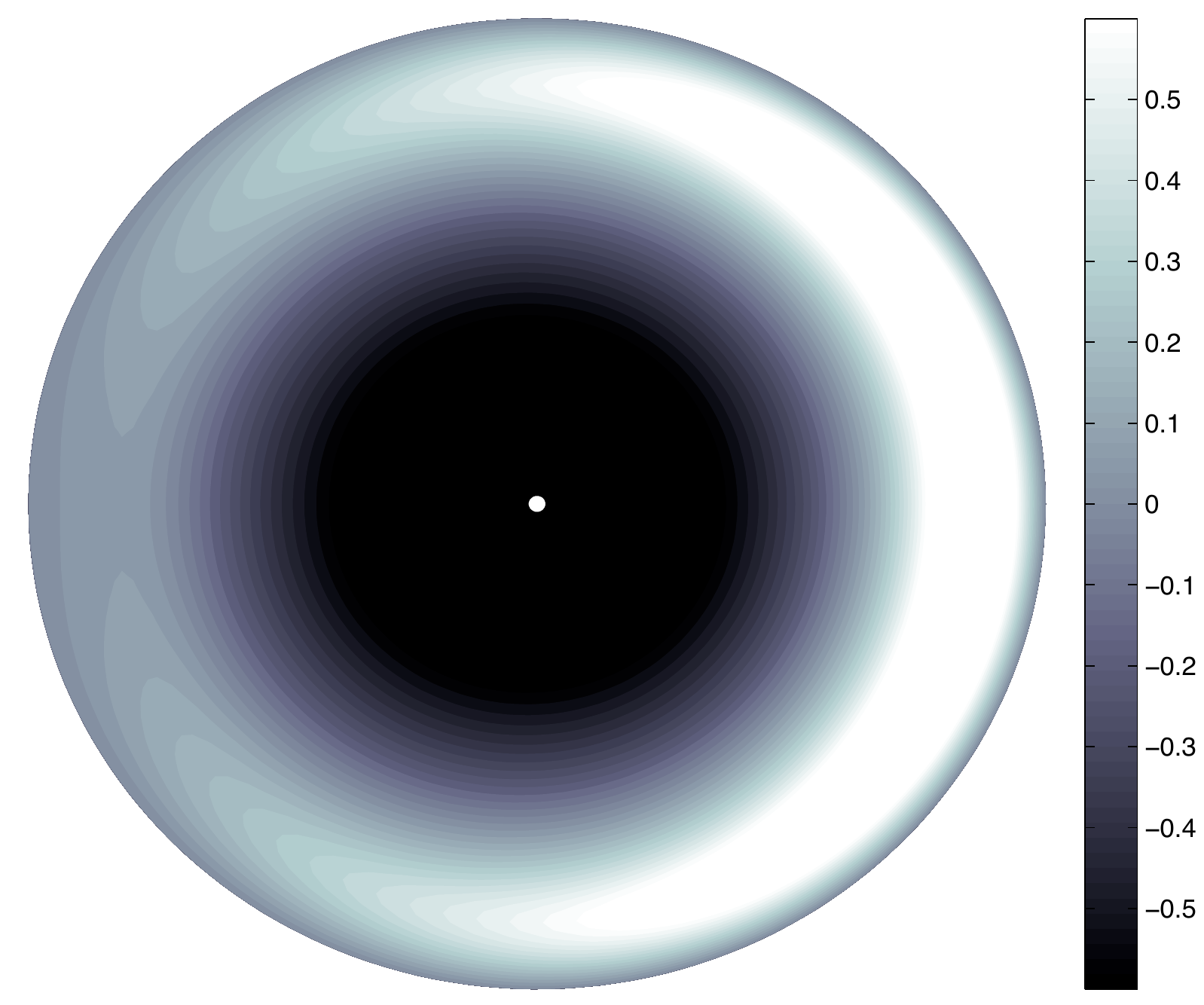}
}
\caption{Model output for deterministic forcing: (a) contours of the streamfunction $\Psi=0.033\,(\eta-3\eta^3+2\eta^4)\sin\phi$, (b) quiver plot of the corresponding in-plane velocities, and (c) contours of the resulting axial velocity field.
 \label{Best}}
\end{figure}

\begin{figure}[!h]
\centering
    \includegraphics[scale=0.35]{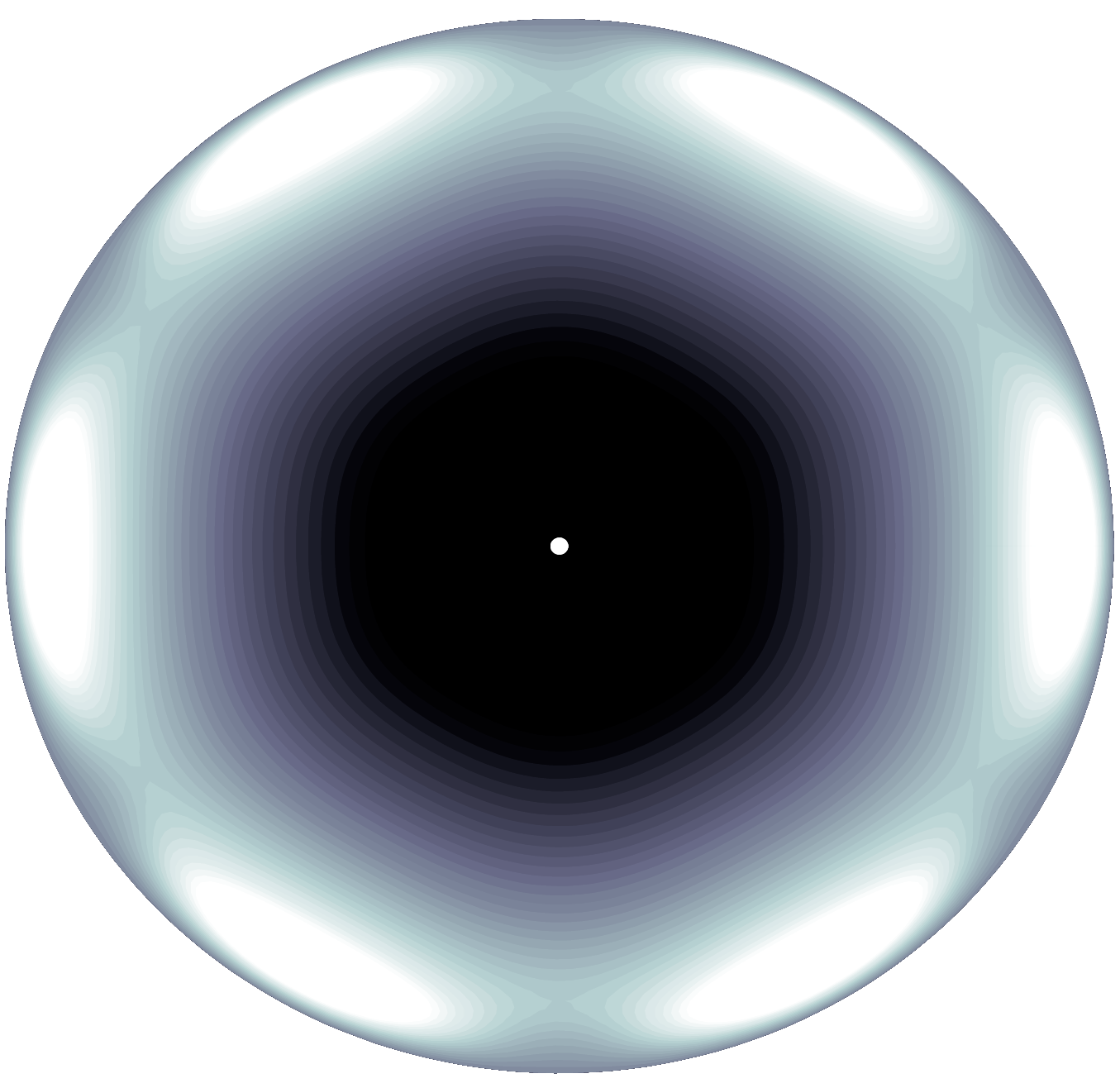}
\caption{Contours of the axial velocity induced by the streamfunction $\Psi(\eta,\phi)=\sin(\pi \eta)^2\sin(6 \phi)$, the light and dark filled contours correspond to regions of the flow respectively faster and slower than laminar.
 \label{Compari}}
\end{figure}

We begin by deriving a simplified version of the 2D/3C model with deterministic forcing to study momentum transfer between the in-plane and axial velocities. The simplified model consists of 3 ordinary differential equations (ODEs) and can be obtained by considering a forcing with only one mode in the azimuthal direction, corresponding to the fundamental mode $n=1$. The choice of the azimuthal dependence of the
forcing is based on the study by \cite{Schmid}, showing that the mode
$(0,1)$ is the most amplified, based on an energy norm. The simplified 2D/3C model predicts the change in mean flow, starting from a laminar base flow, for a given deterministic forcing, and is based on the steady-state equations 
\begin{eqnarray}
\Delta_{1} \Psi_{1}&=&-0.5\, Re\, N(\eta), \\
\Delta_{1} u_{1}&=&0.5\, Re\, \Psi_{1}\, d_{\eta} (U+u_{0}),\\
\Delta_{0} u_{0}&=&-0.5\, Re\, (C \eta - d_{\eta}(\Psi_{1}u_{1})),
\end{eqnarray}
where $N(\eta)$ is the radial shape of the noise, $\Psi=\Psi_{1}\sin\phi$, $u_{x}=u_{0}+u_{1}\cos\phi$, $\Delta_{1}=\eta d_{\eta\eta}+d_{\eta}-\eta^{-1}$, and $\Delta_{0}=\eta d_{\eta\eta}+d_{\eta}$, where $d_{\eta}=\frac{d}{d\eta}$. In order to obtain the simplest model able to capture the blunting of the velocity profile, we can linearize eq. 2 under the assumption of small amplitude forcing. The resulting simplified model contains only one nonlinear term in one ODE, the other two ODEs being linear. The presence of at least one nonlinear term is required to obtain a change in mean flow since
linear models always give the same mean flow as the one used for the
linearization. The model allows us to recover the conclusions of \cite{Ioannou} on the
energy transfer between modes, simply by inspection of the equations,
but this time in terms of momentum transfer and for the pipe instead of Couette flow. The mean turbulent
mode cannot extract momentum from the laminar base flow and is sustained
by the nonlinear interaction between the $(0,1)$ mode of the axial
velocity with the streamfunction, i.e. the $d_{\eta}(\Psi_{1}u_{1})$ term in eq. 3. The non-normality of the system
manifests itself in the linear coupling between the laminar base
flow and the streamfunction which amplifies the disturbances and
generates the axial velocity mode $(0,1)$ by convection of streamwise
momentum (see eq. 2).\\
\\
In order to determine the forcing required to maintain $\Psi_{1}$, we write the streamfunction as a Taylor series at the wall which naturally incorporates the BCs $\Psi_{1}=\frac{d\Psi_{1}}{d\eta}=0$,
\[\Psi_{1}(\eta)=\frac{1}{2!}\frac{d^2\Psi_{1}}{d\eta^2}\Big\vert_{\eta=1}(\eta-1)^2+\frac{1}{3!}\frac{d^3\Psi_{1}}{d\eta^3}\Big\vert_{\eta=1}(\eta-1)^3+...=\alpha_{0} ((\eta-1)^2+\alpha_{3}(\eta-1)^3+...),\]
where $\alpha_{0}$ is the amplitude coefficient and $(\eta-1)^2+\alpha_{3}(\eta-1)^3+...$ is the streamfunction shape,
and compute the corresponding forcing 
\[N(\eta)=-\frac{1}{Re}\left(\partial_{\eta\eta}+\frac{1}{\eta} \partial_{\eta}-\frac{1}{\eta^2}\right)^2 \Psi_{1}(\eta).\]
The parameters $\alpha_{i}$, $i=3,4,...$, must satisfy the regularity condition $\Psi_{1}(0)=0$, i.e. 
\[\sum_{i=3}^{N} \alpha_{i}(-1)^{i}=0,\] 
to ensure continuity at the origin for $\Psi(\eta,\phi)$ when $\phi\rightarrow \phi + \pi$.
The forcing is bounded at the origin if $\Psi_{1}$ has no quadratic term in $\eta$, since $\Delta_{\eta}^2 \,\eta^2=-\frac{3}{\eta^2}$ is unbounded for $\eta=0$.\\
\\
The steady-state equations were solved with $64$ grid points in the radial direction, and at a Reynolds number of 24,600, matching one of the pipe flow experiments of \cite{Toonder}. The streamfunction amplitude coefficient $\alpha_{0}$ was chosen such that the change in mean flow induced by the forcing had the same amplitude at its maxima as in the experiments of \cite{Toonder} at the same Reynolds number. The shape of the streamfunction determined the amount of blunting obtained for a given amplitude coefficient.\\
\\
The near-wall perturbations play a dominant role in the blunting of the velocity profile since the amplification which is proportional to the mean shear and surface area are both maximum at the wall. Hence, streamfunction shapes having a large amplitude near the wall result in a larger blunting of the velocity profile compared to shapes peaking in the core region. Figure \ref{Analytic} illustrates the importance of the near-wall region in obtaining a given amount of blunting of the velocity profile. A fundamental streamfunction $\Psi_{1,a}$ based on the shape $\eta-3\eta^3+2\eta^4$ multiplied by an amplitude coefficient of $0.033$ is selected and compared to streamfunctions obtained by taking the square $\Psi_{1,b}$ and cube $\Psi_{1,c}$ of the fundamental streamfunction shape with amplitude coefficients chosen such that the same amount of blunting is realized in each case. The streamfunctions $\Psi_{1,b}$ and $\Psi_{1,c}$ reach their maximum amplitude respectively of about $0.05$ and $0.25$ in the core of the pipe. Compared to $\Psi_{1,b}$ and $\Psi_{1,c}$, $\Psi_{1,a}$ is relatively flat with a maximum amplitude of $0.017$. Note that even the simple streamfunction $\Psi_{1,a}$ leads to a surprisingly good blunting of the velocity profile.\\
\\
The streamfunction amplitude can be compared to the radial velocity turbulence intensity measured by \cite{Toonder} which is about 1 plus unit or equivalently about 5\% of the bulk velocity. In order to maintain the noise level comparable to the experimental value, we need to choose streamfunctions that have a large amplitude near the wall and are relatively flat over the whole domain. A similar conclusion can be obtained by considering that the blunting results from the advection of axial momentum by the radial velocity so that a large amount of blunting is realized when large radial velocities are present. Taking into account that the radial velocity depends on the azimuthal velocity via the continuity constraint, we need to maximize the ratio $|\frac{u_{r}}{u_{\phi}}|=|\frac{\Psi}{\eta d_{\eta}\Psi}|$. Hence, the flattest streamfunction shape or equivalently the simplest radial dependence results in the largest amplification. In terms of structures, we can imagine that the largest structures corresponding to the modes with the least zero crossings in the radial direction are more able to redistribute momentum over the cross-section of the pipe. The importance of the modes with the least zero crossings is a known feature of turbulent pipe flow: modes with a radial quantum number of 1 in the study of \cite{Duggleby} were shown to capture most of the energy in their dynamical eigenfunction decomposition of turbulent pipe flow. Likewise, the singular modes that are most amplified in the study of \cite{McKeon} exhibit the lowest number of zero crossings.\\
\\
Figure \ref{Compa} illustrates the difference between the polynomial streamfunction $\Psi=0.033\,(\eta-3\eta^3+2\eta^4)\sin\phi$ and a trigonometric streamfunction whose series expansion contains an infinite number of terms. The two streamfunctions yield the same velocity profile but the trigonometric streamfunction has a slightly larger amplitude further from the wall since near the wall the presence of higher order terms in the series expansion decreases the streamfunction amplitude and hence the overall amplification.\\
\\
Based on our numerical study, the streamfunction $\Psi=0.033\,(\eta-3\eta^3+2\eta^4)\sin\phi$ leads to a blunting of the velocity profile whose maximum amplitude matches the experimental data at the same Reynolds number while maintaining a noise level over the pipe cross-section that is consistent with the experimentally measured rms amplitude of the turbulence fluctuations. The streamfunction $\Psi=0.033\,(\eta-3\eta^3+2\eta^4)\sin\phi$ corresponds to a noise profile $N(\eta)=-90/Re$ that is constant in the radial direction and is considered to be the most likely to be realized in a flow, since it yields the largest amplification and exhibits the simplest radial dependence. By simplest radial dependence we mean the radial profile with the least zero crossings.\\
\\
The results obtained with the simplified 2D/3C model also show that the velocity profile is relatively independent of the
radial shape of the forcing and streamfunction, i.e. the profile is
robust, see Figures \ref{Analytic} and \ref{Compa}.  Note that the maxima of the velocity profiles are situated further from the wall, compared to experimental data obtained
by \cite{Toonder}, as can be seen on Figure \ref{Compa}, which likely corresponds to the neglect of the influence of the small scales near the wall by the streamwise-constant model.\\
\\
The contours of the streamfunction $\Psi=0.033\,(\eta-3\eta^3+2\eta^4)\sin\phi$ are plotted on Figure \ref{Best} together with a quiver plot of the corresponding in-plane velocities and the resulting contours of the axial velocity field. The streamfunction exhibits two counter-rotating rolls which advect the mean shear to create a low- and a high-speed streak of axial velocity defined with respect to the laminar base flow. The high-speed streak sits near the wall whereas the low-speed streak is localized near the centerline. Hence, the flow is on average faster near the wall and slower at the center as is the case for the velocity profile of turbulent pipe flow. The amplification factor between the in-plane and streamwise velocities, defined as the ratio of the extrema, is about $100$ in this case. When the
azimuthal wavenumber is chosen to match the azimuthal dependence of the spatial puffs observed by \cite{Hof}, the velocity fields produced by our model are remarkably similar to the
velocity field near the trailing edge of the puffs, i.e. the wall-normal position of the high- and low-speed streaks is accurately captured by the model as can be seen in Figure \ref{Compari} compared to Figure 2 (E),(F) in the paper by \cite{Hof}. Moreover the model captures the merging of the low-speed streaks and their congregation near the center of the pipe which is observed in the experiments of \cite{Hof} but is not present in the travelling wave (TW) solutions of the NSE.
\\
\section{Stochastic forcing of the 2D/3C model}
\begin{figure}[!h]
\centering
\subfigure[] 
{
    \label{fig:sub:a}
    \includegraphics[scale=0.45]{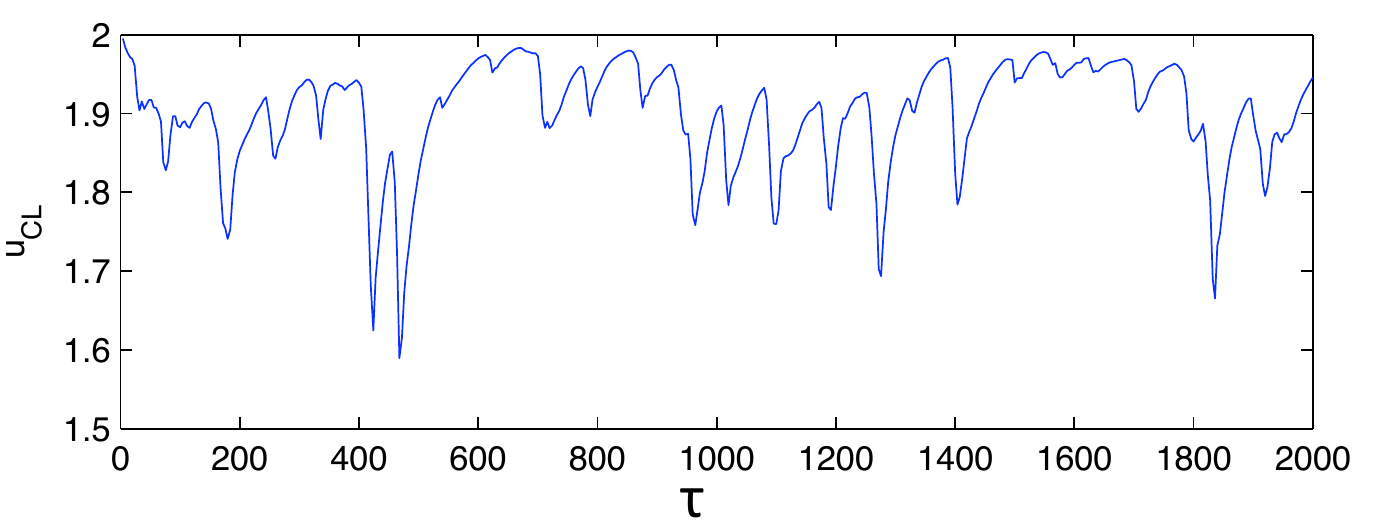}
}
\subfigure[]
{
    \label{fig:sub:b}
    \includegraphics[scale=0.45]{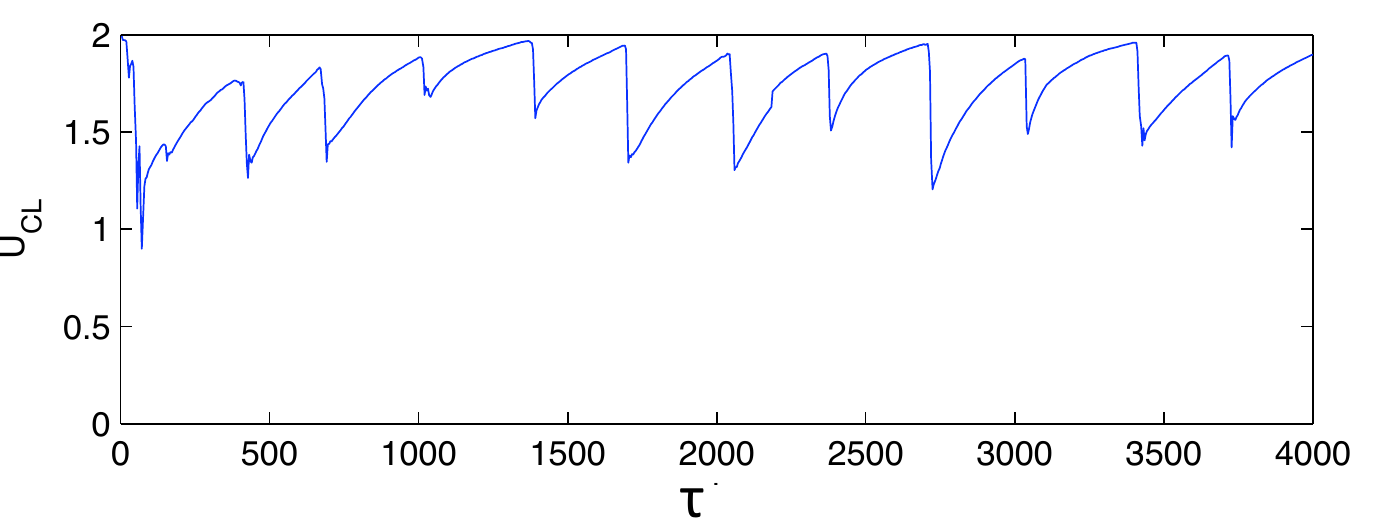}
}
\subfigure[]
{
    \label{fig:sub:c}
    \includegraphics[scale=0.45]{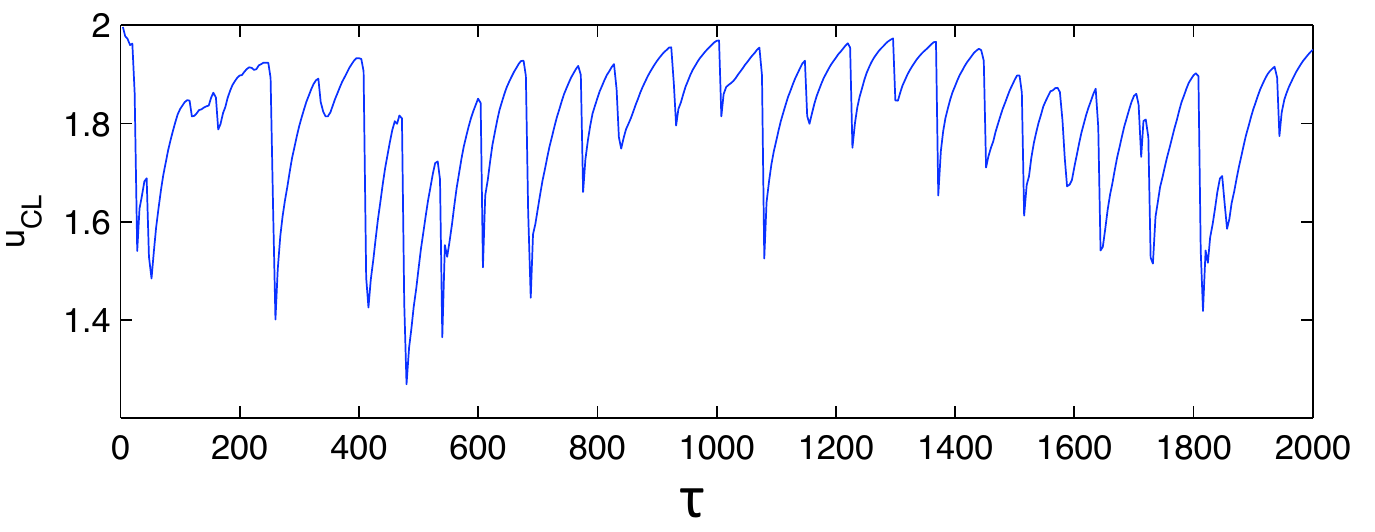}
}
\subfigure[]
{
    \label{fig:sub:d}
    \includegraphics[scale=0.45]{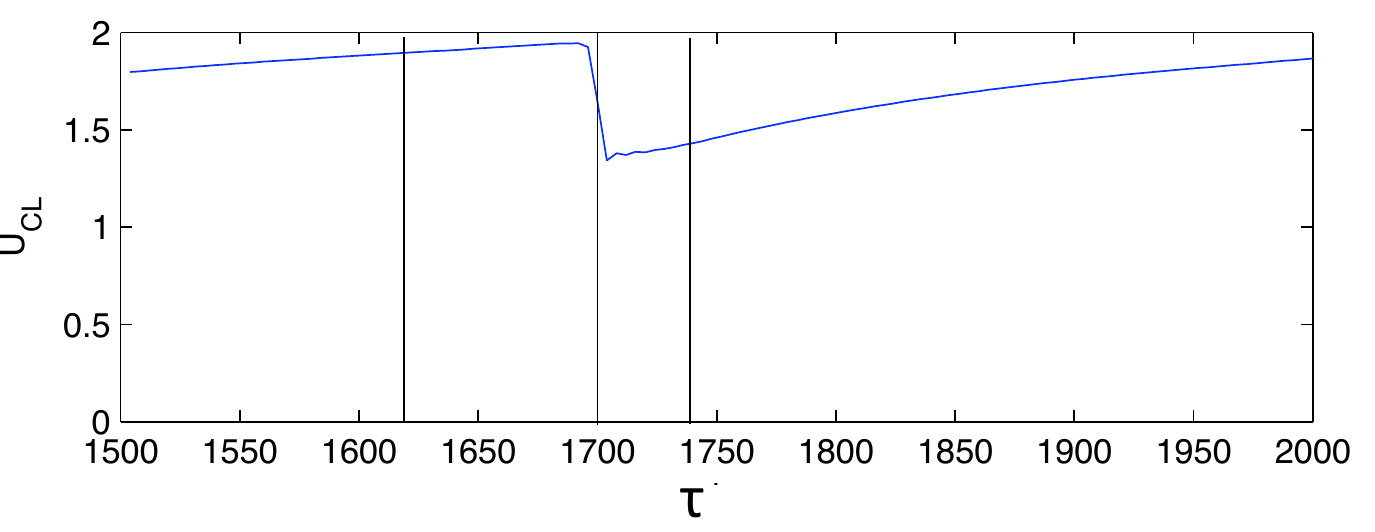}
}
\caption{Time traces of the centerline velocity from three different simulations respectively at $Re=2,200$ with $0.0005$ and $0.002$ rms noise levels (a), (c) and at $Re=10,000$ with $0.002$ rms noise level (b). The resolution in the radial direction is $N=48$. (d) Zoom on the time
interval during which the samples of Figure \ref{movie} are taken. The vertical
lines indicate the sampling instants.
 \label{traces}}
\end{figure}

\begin{figure}[!h]
\centering
\subfigure[] 
{
    \label{fig:sub:a}
    \includegraphics[scale=0.35]{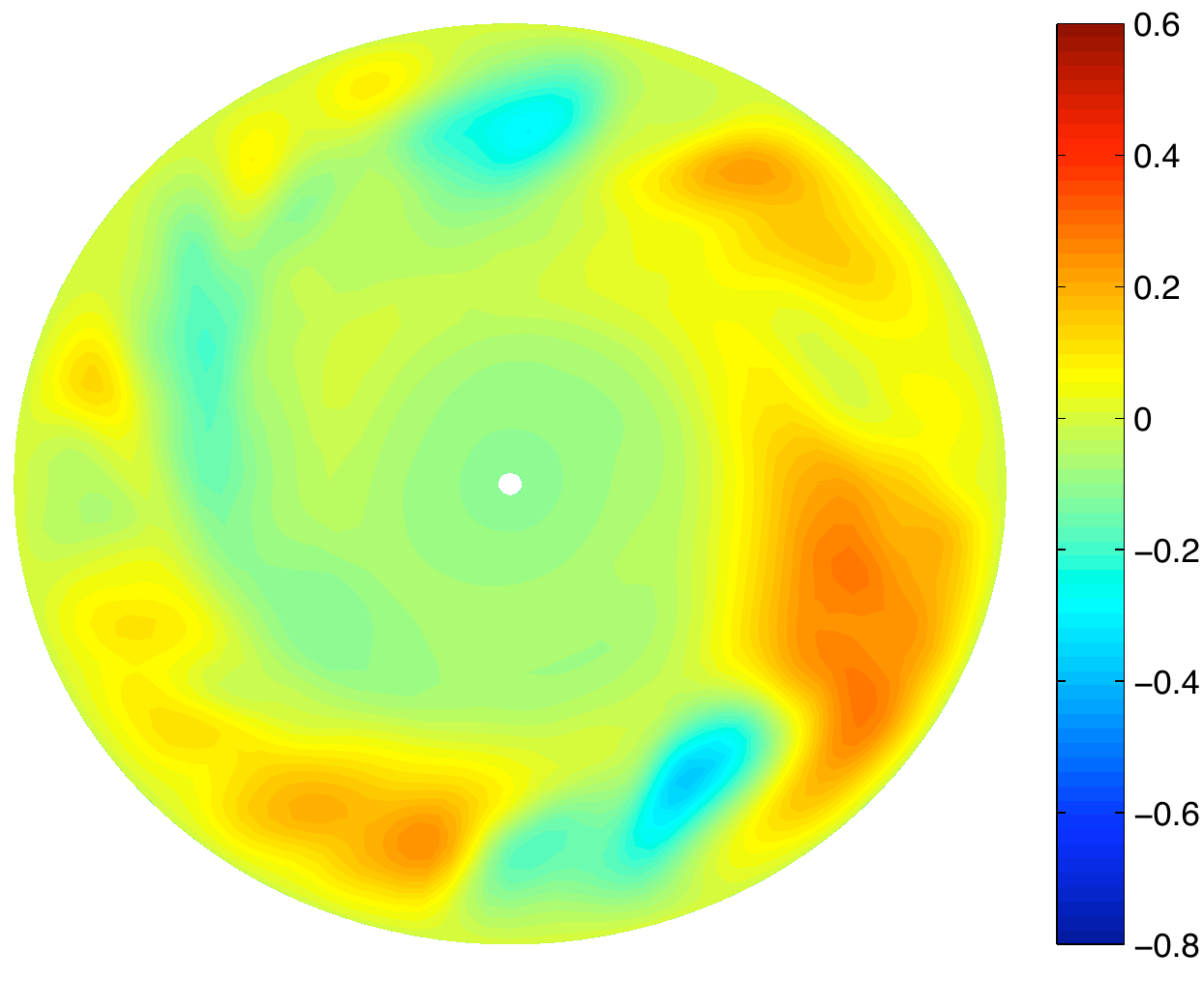}
}
\subfigure[]
{
    \label{fig:sub:b}
    \includegraphics[scale=0.35]{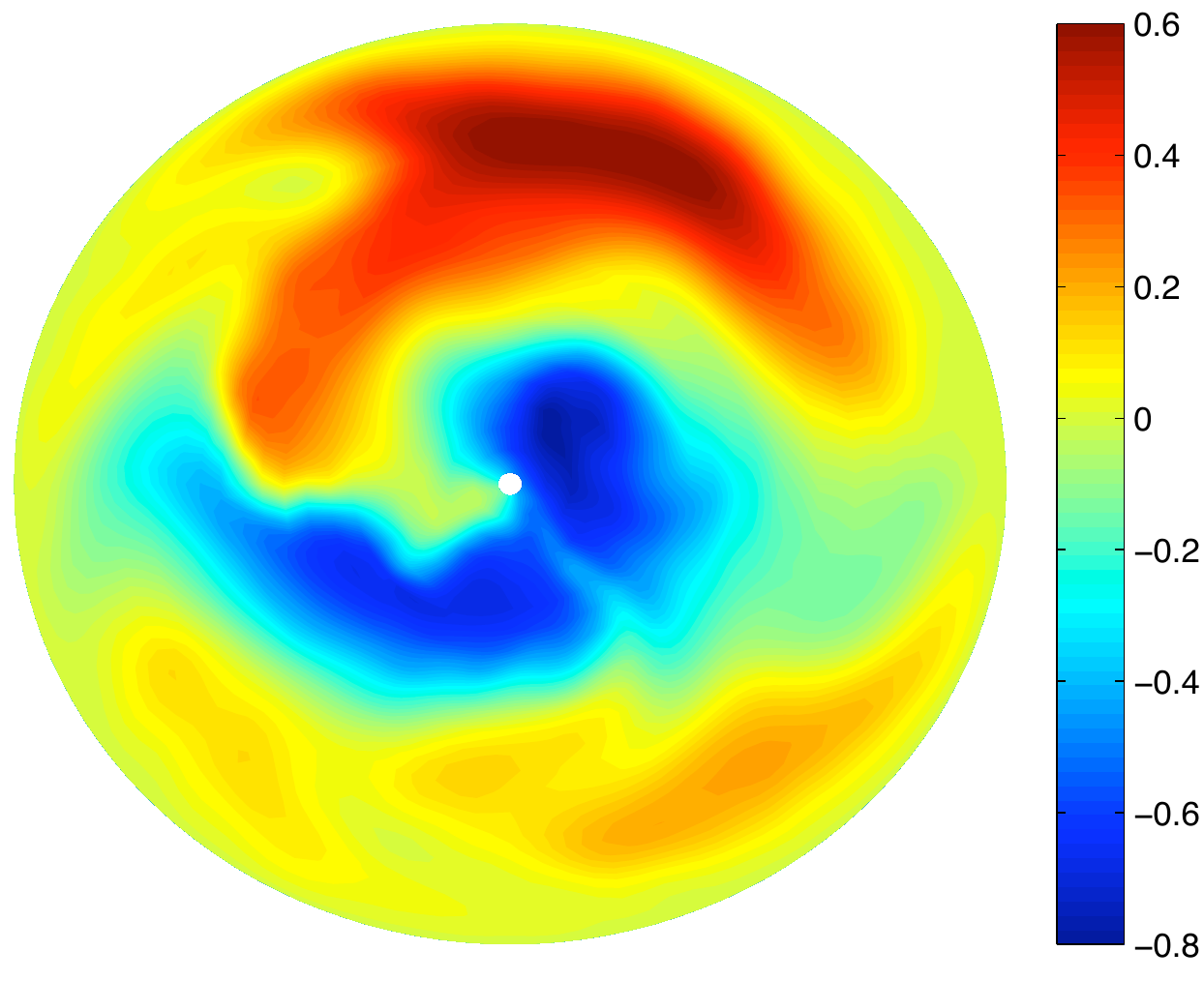}
}
\subfigure[]
{
    \label{fig:sub:c}
    \includegraphics[scale=0.35]{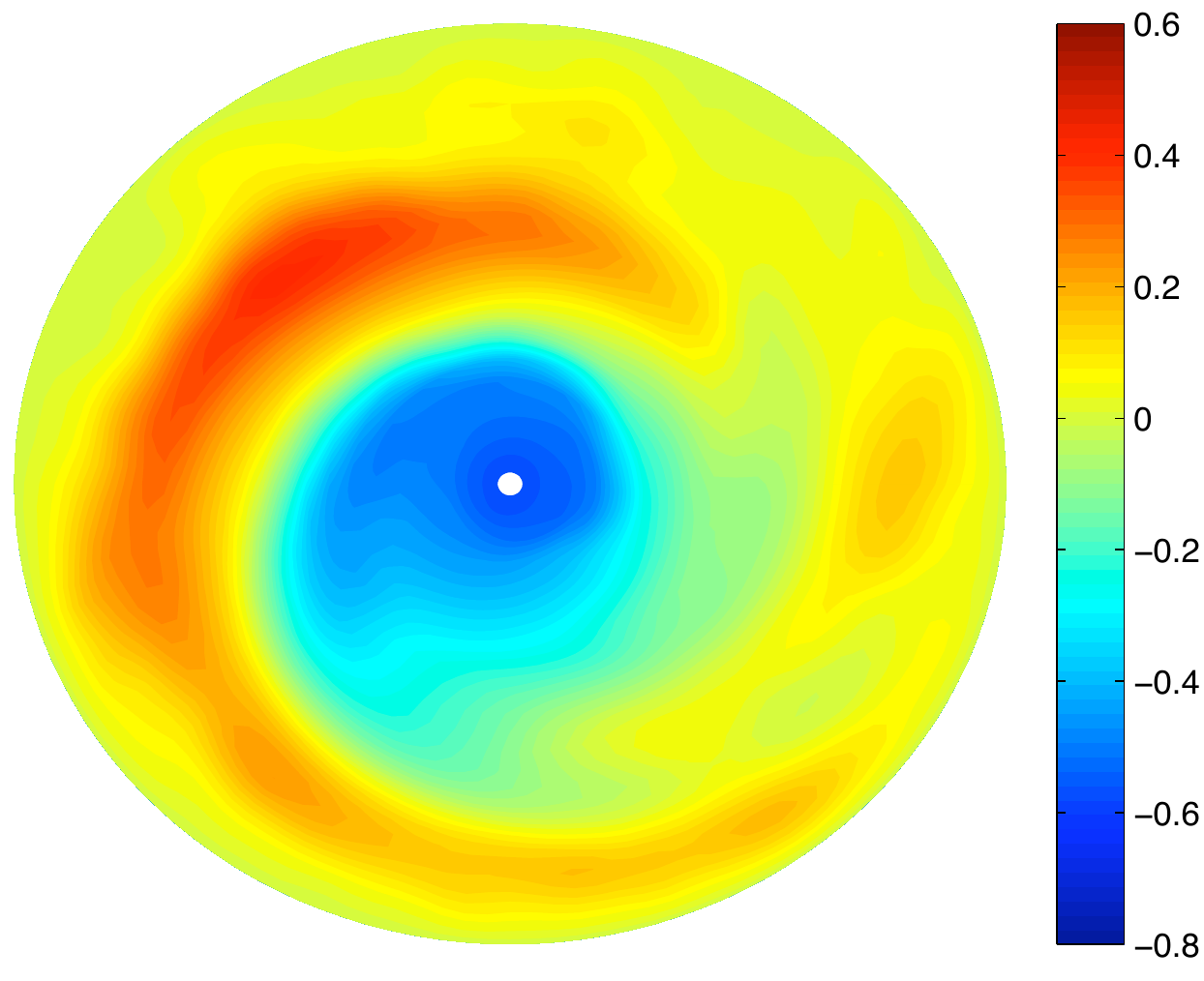}
}
\subfigure[]
{
    \label{fig:sub:d}
    \includegraphics[scale=0.35]{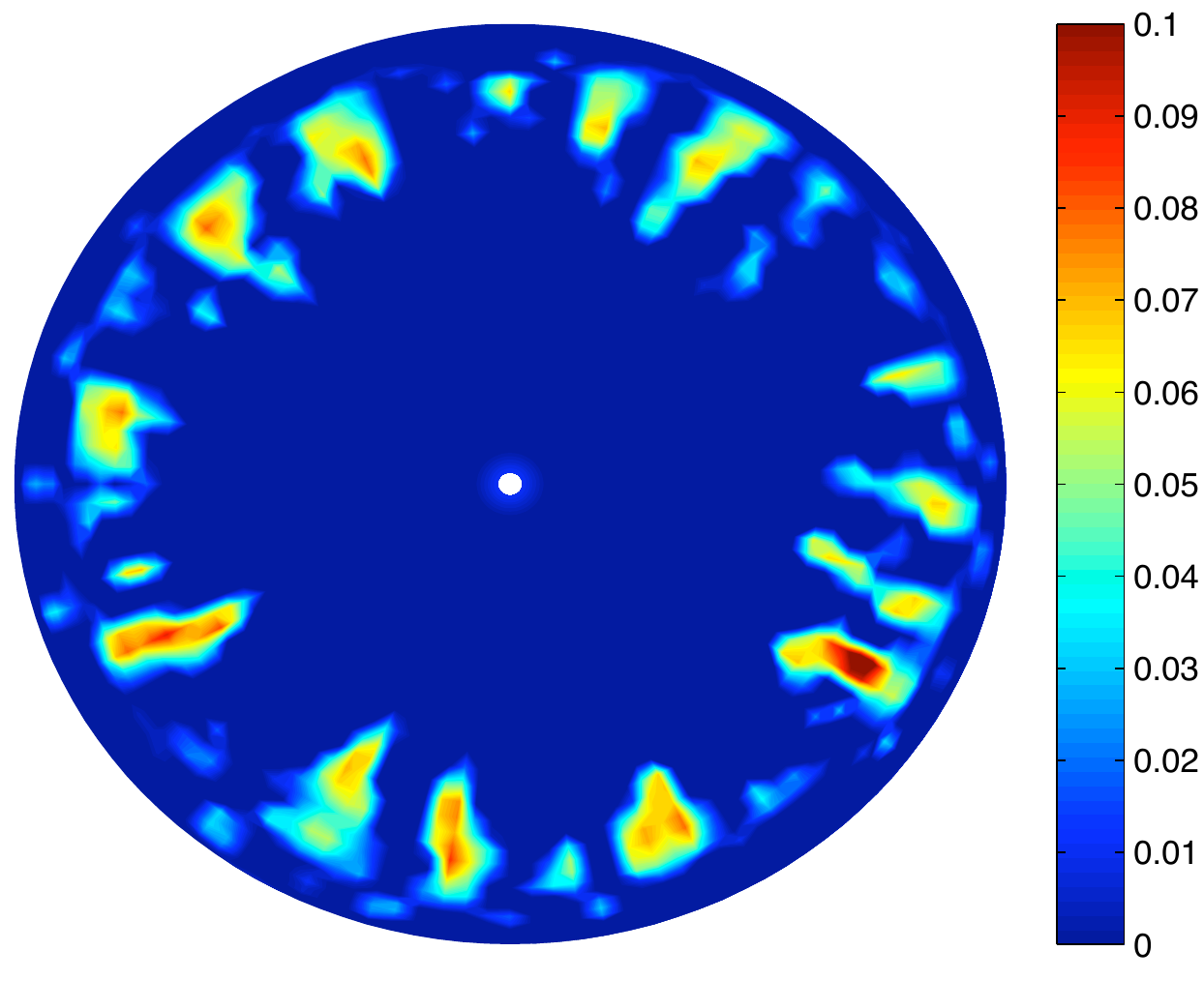}
}
\subfigure[]
{
    \label{fig:sub:e}
    \includegraphics[scale=0.35]{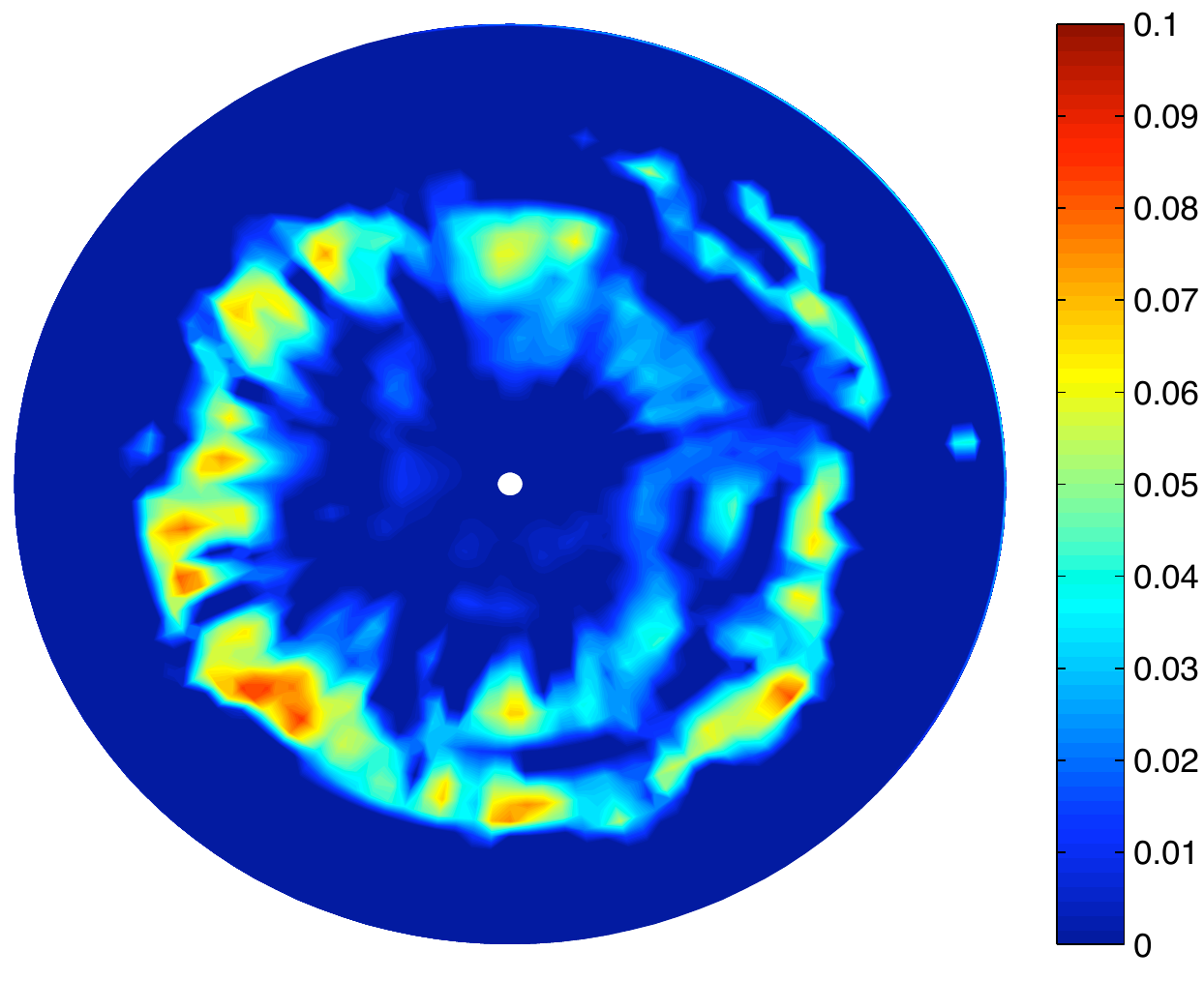}
}
\subfigure[]
{
    \label{fig:sub:f}
    \includegraphics[scale=0.35]{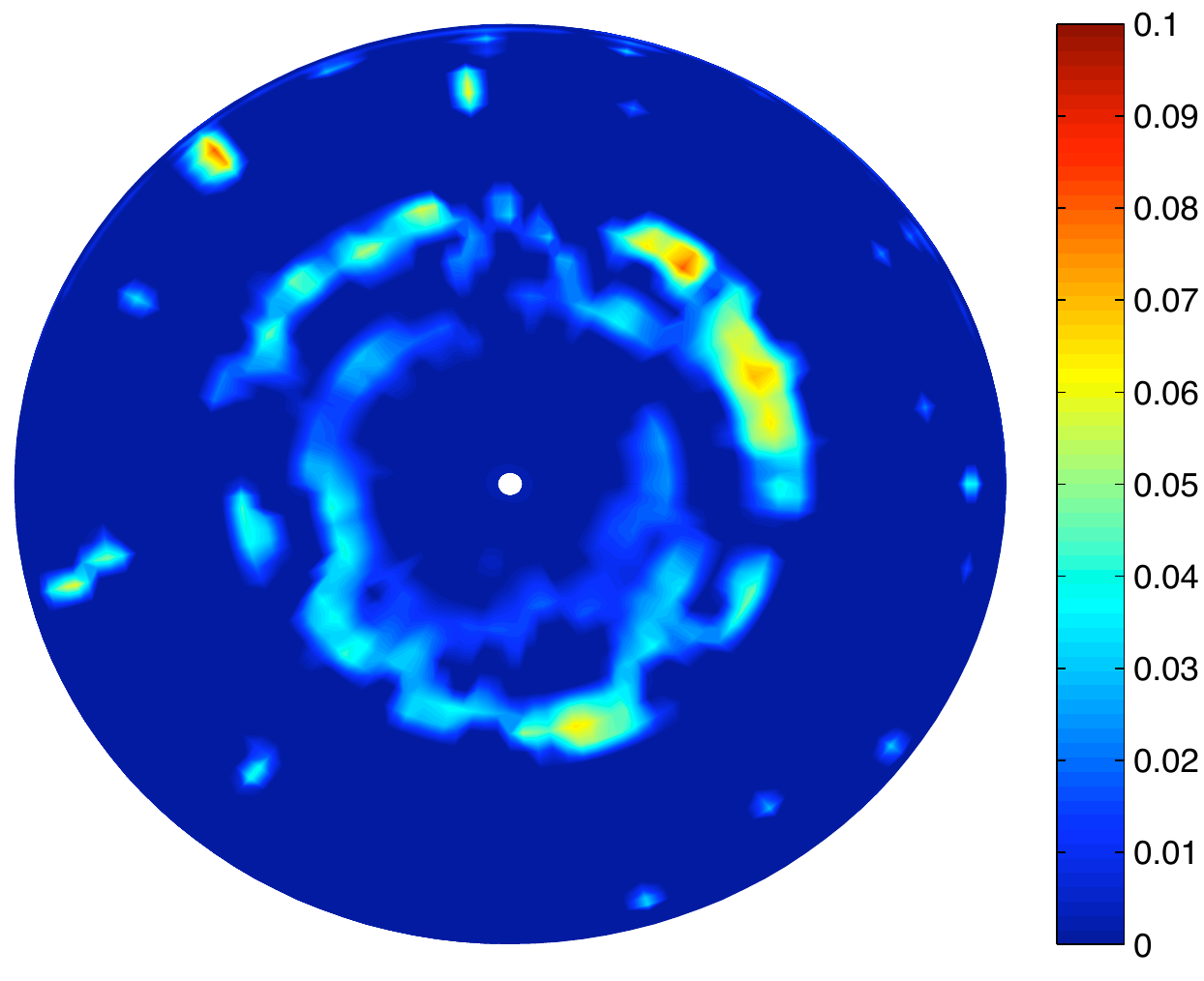}
}
\caption{Contours of the axial velocity, subfigures (a) to (c), and of
  the swirling strength for the in-plane velocities, subfigures (d) to (f), computed respectively at $t=1620$, $t=1700$, and
  $t=1740$ dimensionless time units. The swirling strength is defined as the magnitude of the imaginary part of the in-plane velocity gradient eigenvalues.
 \label{movie}}
\end{figure}

\begin{figure}[!h]
\centering
    \includegraphics[scale=0.35]{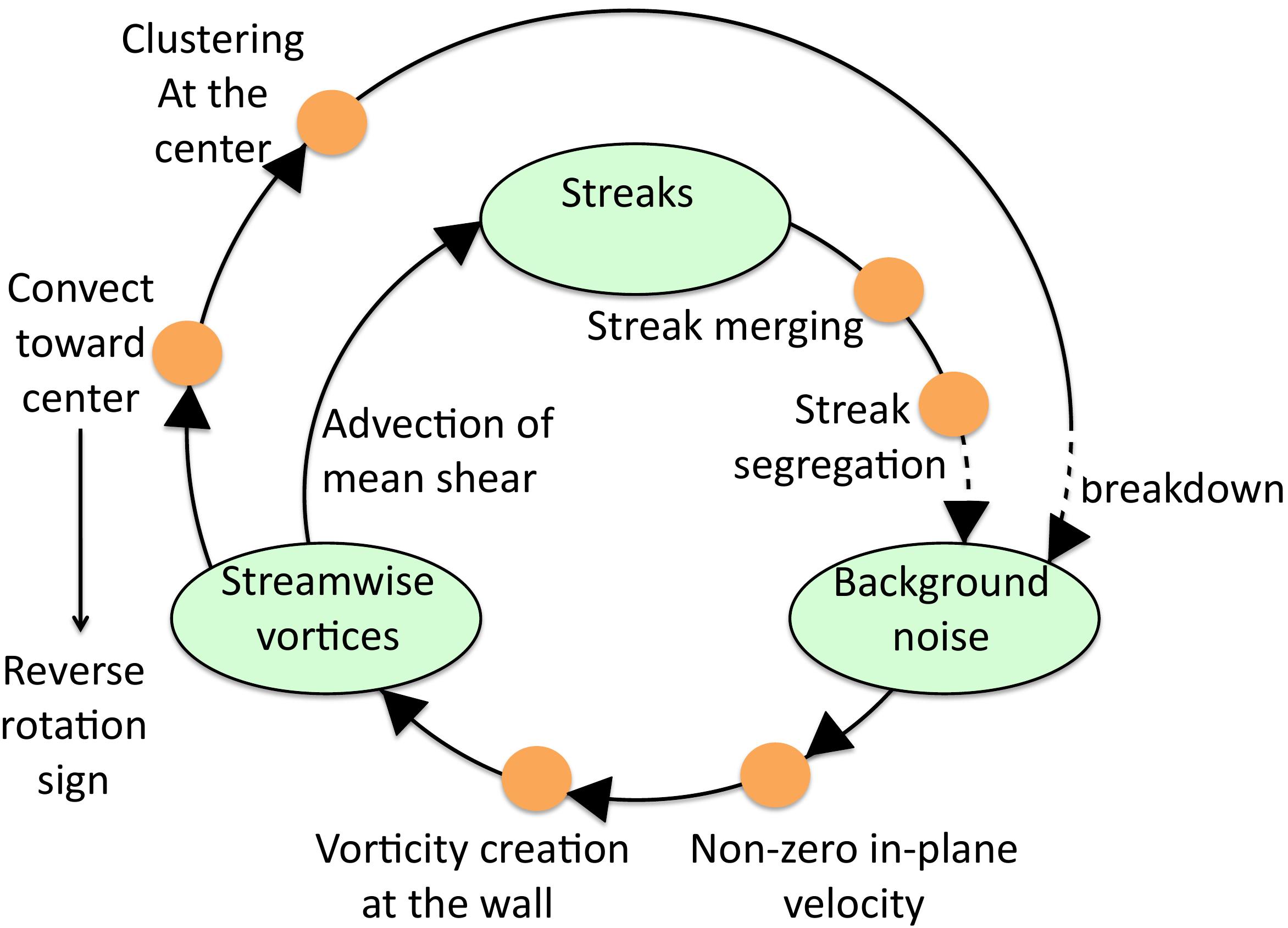}
\caption{Diagram detailing the different stages of the SSP exhibited by our model. The dashed lines represent unmodeled effects. As the streamwise vortices convect toward the center of the pipe the sign of the azimuthal velocity is reversed such that the rotation sign changes from one cycle to the next. 
 \label{SSP}}
\end{figure}

We now simulate the 2D/3C model forced by small amplitude white noise which takes into account unmodeled effects and perturbations that are always present in experiments. Stochastic forcing has the advantage over deterministic forcing of not relying on any assumption regarding the spatial and temporal dependence of the perturbations and the resulting velocity field will reflect the direction of maximum amplification. The noise $\mathcal{N}_{\psi}$ is applied at each grid point in space and at every time step.\\ 
\\
The model shows that the presence of background noise in the pipe cross-section results in the formation of streamwise-constant vortices which advect axial momentum to create high- and low-speed streaks. The vortices interact nonlinearly with the streaks to segregate them, i.e. to convect the low-speed streaks toward the center and the high-speed streaks toward the wall, leading to the blunting of the velocity profile. Hence, we demonstrate that the blunting of the velocity profile is essentially a nonlinear two-dimensional phenomenon. The time evolution of the flow field is characterized by the quasi-periodic generation of  ``streamwise-constant puffs'' followed by their decay and the return of the flow close to the laminar state, i.e. each bursting event is followed by quiescent flow equivalent to the laminar regions that separate the puffs in the experiments of \cite{Lindgren}. The generation and subsequent decay of the puffs is based on a very simple self-sustaining mechanism driven by the background noise.\\ 
\\
The time traces of the centerline velocity for two different Reynolds numbers are reproduced on Figure \ref{traces} and show numerous sharp drops which we identify as the signature of ``streamwise-constant puffs'' before increasing smoothly nearly back to its laminar value. Since there is no grid point at the centerline, we approximate the centerline velocity by averaging the axial velocity in the azimuthal direction over the grid points closest to the center of the pipe. The signatures of the ``streamwise-constant puffs"  are remarkably similar to the spatial evolution of the centerline velocity from the trailing to the leading edge of the spatial puffs in the numerical simulations of \cite{Shimizu}. We define a puff generation timescale as the time elapsed between two sharp drops of the centerline velocity. At $Re=2,200$, the timescale is about $75$ dimensionless time units based on the pipe radius and compares well with the time scale of the puffs reported in the literature \citep{Nishi}, e.g. a puff of length $20D$ convected at nearly the bulk velocity and which is separated from the next puff by a laminar region of length equivalent to one puff would lead to a dimensionless timescale of $80$.\\
\\
The puff generation timescale is an increasing function of the Reynolds number, reflecting the fact that puffs tend to be longer in experiments as the Reynolds number increases,
reaching $330$ at $Re=10,000$, but is relatively independent of the noise amplitude. The drop in centerline velocity associated with the trailing edge of a puff is sharper and stronger for larger forcing amplitudes. The periodicity of the puff signatures in the centerline velocity time traces comes from the SSP driven by the background noise which quasi-periodically generates puffs separated in time by a return of the flow close to the laminar state.\\
\\ 
The SSP starts with the formation of streamwise
vortices inside the domain, due to the stochastic forcing, and at the wall,
as a response to the forcing, in order to enforce the no-slip boundary
condition. The vorticity at the wall has the same order of magnitude
as inside the domain but plays a more important role in the flow dynamics
since the amplification and surface area are maximum at the
wall. The importance of the near-wall vorticity at the beginning of the SSP is in agreement with the simulations of \cite{vanDoorne} in which the vorticity field near the trailing edge of a puff is dominated by near-wall quasi-streamwise vortices. The three main stages in the evolution of a ``streamwise-constant puff''
are plotted on Figure \ref{movie} \textbf{(a)} to \textbf{(f)}. During the first stage, the near-wall
streamwise vortices start to move toward the center of the pipe
\textbf{\ref{movie}(d),(e)} and create streaks by convecting the axial momentum
\textbf{\ref{movie}(a)}. The radial motion of the streamwise vortices corresponds to a lift up of the vortices away from the wall as observed in the simulations of \cite{vanDoorne} if we consider the evolution of the vorticity field projected in a plane moving at the bulk velocity. The second stage consists of the segregation of
the high- and low-speed streaks, the latter being convected toward the
center of the pipe, resulting in the blunting of the velocity profile characteristic of turbulent pipe flow \textbf{\ref{movie}(b)}. Once a low-speed streak reaches the
center, the centerline velocity drops sharply, as can be
seen on the time traces on Figure \ref{traces} (b). 
Finally the streamwise vortices and streaks decay \textbf{\ref{movie}(c),(f)} and the flow returns
close to the laminar state before the next cycle starts. The different stages of the SSP exhibited by our model are summarized in the diagram shown on Figure \ref{SSP}.\\
\\
The ``streamwise-constant puffs'' are composed of streamwise vortices and streaks and the time evolution of the velocity field is remarkably similar to the flow visualizations by \cite{Hof} in transitioning pipe flow when a puff is observed in a reference frame moving with the bulk velocity. The segregation of the high- and low-speed streaks observed in the experiments is accurately captured by the model as well as the streak merging. Streak merging in experiments was reported by \cite{Hof} who showed that the number of streaks in the cross-section decreases due to their merging as we move from the trailing edge to the leading edge of a puff. The time evolution of the vorticity field in our simulations is consistent with the projection of the 3-dimensional vorticity field inside a puff computed by \cite{vanDoorne} in a plane moving at the bulk velocity and is characterized by a lift up of the streamwise vortices away from the wall as we move from the trailing to the leading edge of the puff. \\ 
\\
Note that if we relax the BCs to allow for slip in the azimuthal direction, i.e. we use a shear-stress free condition, the simulations capture the creation of streamwise vortices and streaks as well as the blunting of the velocity profile, but we do not observe clearly the cyclic generation of puffs in the time evolution of the full velocity field or their signature in the time traces of the centerline velocity. However, for a given forcing amplitude, the amount of blunting realized with the slip BC in the azimuthal direction is larger than with no-slip.\\
\\
The SSP we have just described is significantly simpler than the ones invoked in the literature but does capture the minimum turbulence dynamics and
produces flow fields dominated by streamwise vortices and streaks whose temporal evolution and topology compares well to experimental visualizations and numerical simulations of puffs projected in a plane convecting at the bulk velocity. The SSP is similar to the one described in \cite{Shimizu}, except
that the streamwise vortices are generated here as a consequence of the
noise forcing. Hence, our model shows that using white noise forcing to generate the streamwise vortices is sufficient to capture the formation of the streaks and their segregation resulting in a blunting of the velocity profile and that the overall flow dynamics are relatively insensitive to the particular regeneration process invoked to produce the streamwise vortices. In addition, our simulations are significantly less computationally intensive than in the works of \cite{Shimizu} and \cite{vanDoorne} since our domain is $2D$ and we do not need to track the position of the puffs in time.\\
\\
Interestingly, our simulations of the stochastically forced 2D/3C model for pipe flow do not reach a statistically steady state, regardless of the amount of forcing and Reynolds number, even though the same model applied to Couette flow and described in \cite{Dennice} converged to a statistically steady state with a velocity profile in good agreement with the profiles from full 3D simulations at the same Reynolds number, provided that the amount of forcing is appropriately chosen. Hence, three-dimensional effects are necessary to reach fully developed turbulence in the pipe, i.e. to allow for the puffs to destablize and form slugs which will expand to cover the whole flow domain. The absence of convergence of our model can be linked to the more complex dynamics present in the pipe compared to Couette flow and to the fact that the mean shear of the base flow is no longer constant over the cross-section.\\
\\
The presence of background noise in our model can also
be justified by considering that the streamwise streaks are highly
unstable, see \cite{Meseguer}, and therefore would break down in a 3D flow and
increase the amount of background disturbances. The radial shape of the noise is directly related to the amount of swirl present in the
simulations and therefore suggests a method to control the flow by shaping the noise forcing in the spirit of the simple control mechanism developed by \cite{Hof2} which reduces the inflection points of the velocity profile and leads to a relaminarization of the flow.\\ 
\\ 
\newpage
\section{Conclusions}
We presented a streamwise-constant, 2D/3C model of turbulent pipe flow which captures the main features associated with transition to turbulence, i.e. the blunting of the velocity profile and the generation of streamwise vortices and streaks, and is significantly more tractable than the full Navier-Stokes equations. Using a time-invariant deterministic forcing, we showed that the velocity profile is robust with respect to variations in the forcing profile and we produced realistic velocity fields that are remarkably similar to the flow visualizations by \cite{Hof} near the trailing edge of a puff.\\
\\
Thanks to the significant reduction in complexity achieved by our model compared to the full Navier-Stokes equations, we were able to study momentum transfers between the mean and perturbation modes directly from the governing equations with a deterministic streamfunction input. We concluded that momentum is extracted from the laminar base flow by the perturbations via linear nonnormal mechanisms followed by the nonlinear interactions of the perturbations which result in a change in mean flow. Hence, our model allowed us to isolate the basic mechanisms leading to the blunting of the velocity profile in pipe flow.\\   
\\
Under stochastic forcing, the model generates ``streamwise-constant puffs'' at a frequency that depends on the Reynolds number but not on the forcing amplitude. Contrarily to Couette flow, the 2D/3C pipe flow simulations do not reach a steady state, implying that for the pipe some three-dimensional effects are necessary in order to reach fully developed turbulence. However, the time evolution of the velocity fields produced by our simulations is remarkably similar to the flow visualizations by \cite{Hof} in transitioning pipe flow when a puff is observed in a reference frame moving at the bulk velocity. The segregation of the high- and low-speed streaks observed in the experiments is accurately captured by the model as well as the streak merging and the lift up of the streamwise vortices away from the wall. We showed that the dynamics governing the generation of puffs in pipe flow transition can be captured by a streamwise-constant model and are relatively insensitive to the particular regeneration mechanisms invoked to produce the streamwise vortices. Even white noise forcing is sufficient to regenerate the vortices.\\
\\  
The authors gratefully acknowledge the support of the AFOSR grant FA 9550-09-1-0701 (program
  manager John Schmisseur).\\
\\ 
\bibliography{2d3cbib}
\bibliographystyle{jfm}
\end{document}